\documentclass[12pt]{article}
\usepackage{amsmath}
\usepackage{amssymb}
\usepackage{amsthm}

\begin{document}

\begin{center}
{\large{\bf Bosonization in the Path Integral Formulation}}
\end{center}
\vskip .5 truecm
\centerline{\bf Kazuo Fujikawa$^1$ and Hiroshi Suzuki$^2$}
\vskip .4 truecm
\centerline{\it $^1$ Mathematical Physics Laboratory,}
\centerline{\it RIKEN Nishina Center, Wako 351-0198, Japan}
\vspace{0.3cm}
\centerline{\it $^2$ Department of Physics,}
\centerline{\it Kyushu University, 6-10-1 Hakozaki, Higashi-ku, Fukuoka,
812-8581, Japan}
\vskip 0.5 truecm

\begin{abstract}
We establish the direct $d=2$ on-shell bosonization
$\psi_{L}(x_{+})=e^{i\xi(x_{+})}$ and~$\psi_{R}^{\dagger}(x_{-})=e^{i\xi(x_{-})}$ in
path integral formulation by deriving the off-shell relations
$\psi_{L}(x)\psi_{R}^{\dagger}(x)=\exp[i\xi(x)]$
and~$\psi_{R}(x)\psi_{L}^{\dagger}(x)=\exp[-i\xi(x)]$. Similarly, the on-shell
bosonization of the bosonic commuting spinor,
$\phi_{L}(x_{+})=ie^{-i\xi(x_{+})}\partial^{+}e^{-i\chi(x_{+})}$,
$\phi^{\dagger}_{R}(x_{-})=e^{-i\xi(x_{-})-i\chi(x_{-})}$
and~$\phi_{R}(x_{-})=ie^{i\xi(x_{-})}\partial^{-}e^{+i\chi(x_{-})}$,
$\phi^{\dagger}_{L}(x_{+})=e^{i\xi(x_{+})+i\chi(x_{+})}$,
is established in path integral formulation by deriving the off-shell relations
$\phi_{L}(x)\phi^{\dagger}_{R}(x)=ie^{-i\xi(x)}\partial^{+}e^{-i\chi(x)}$
and~$\phi_{R}(x)\phi^{\dagger}_{L}(x)=ie^{i\xi(x)}\partial^{-}e^{i\chi(x)}$.
\end{abstract}

\makeatletter


\section{Introduction}
It is known that the dynamical degrees of freedom of spinors are consistently
described by the bosonic degrees of freedom in $d=2$~dimensional
space-time~\cite{Coleman:1974bu,Mandelstam:1975hb}. This fact, which is called
the bosonization of spinors, has been analyzed in detail and it has been
applied to various fields of theoretical physics such as conformal field theory
and string theory and also to condensed matter physics~\cite{Polyakov:1983tt,%
Polyakov:1984et,Witten:1983ar,Gonzales:1984zw,DiVecchia:1984df,%
Polchinski:1998rr,peter,Abdalla:1991vua,Stone:1995ys}. The path integral
formulation of bosonization, namely, a systematic derivation of bosonization
rules in the framework of path integrals has been initiated at the early stage
of this subject~\cite{Roskies:1980jh,GamboaSaravi:1981zd,Furuya:1982fh,%
Naon:1984zp,Fujikawa:2004cx}. The path integral bosonization has been applied
to various problems~\cite{Banerjee:1984kc}--\cite{
Bufalo:2014uza}, but, to our
knowledge, the simplest bosonization rule in the form~$\psi(x)\sim e^{i\xi(x)}$
appears to be missing. In this paper, we show how to realize this direct
bosonization in path integral formulation.

We first recapitulate the known path integral bosonization rule
\begin{align}
   e^{iW(v_{\mu})}
   &=\int\mathcal{D}\Bar{\psi}\,\mathcal{D}\psi\,
   \exp\left\{i\left(\frac{1}{2\pi}\right)
   \int d^{2}x\,\left[
   \Bar{\psi}i\gamma^{\mu}\partial_{\mu}\psi+v_{\mu}\Bar{\psi}\gamma^{\mu}\psi
   \right]\right\}
\notag\\
   &=\int\mathcal{D}\xi\,
   \exp\left\{i\left(\frac{1}{4\pi}\right)
   \int d^{2}x\,
   \left[
   \frac{1}{2}\partial^{\mu}\xi(x)\partial_{\mu}\xi(x)
   -2v_{\mu}\epsilon^{\mu\nu}\partial_{\nu}\xi(x)
   \right]\right\},
\label{eq:(1)}
\end{align}
where the normalization of field variables is chosen to make the bosonization
rule simpler. The basic ingredient to derive this path integral bosonization
rule is the chiral Jacobian~\cite{Fujikawa:1979ay}
\begin{equation}
   \ln J(\beta)
   =\frac{i}{\pi}\int d^{2}x\,
   \beta(x)\left[
   \partial^{\mu}A_{\mu}+\frac{1}{2}\epsilon^{\mu\nu}
   \left(\partial_{\mu}V_{\nu}-\partial_{\nu}V_{\mu}\right)\right]
\label{eq:(2)}
\end{equation}
for an infinitesimal chiral transformation
$\psi(x)\to e^{i\beta(x)\gamma_{5}}\psi(x)$
and~$\Bar{\psi}(x)\to\Bar{\psi}(x)e^{i\beta(x)\gamma_{5}}$ in the generic path
integral
\begin{equation}
   \int\mathcal{D}\Bar{\psi}\,\mathcal{D}\psi\,
   \exp\left\{i\left(\frac{1}{2\pi}\right)
   \int d^{2}x\,\left[
   \Bar{\psi}i\gamma^{\mu}
   \left(\partial_{\mu}-iV_{\mu}-iA_{\mu}\gamma_{5}\right)\psi\right]\right\}.
\label{eq:(3)}
\end{equation}
To use this Jacobian factor, one may replace
$v_{\mu}=\partial_{\mu}\alpha(x)+\epsilon_{\mu\nu}\partial^{\nu}\beta(x)$, which
is valid in~$d=2$, in the fermionic path integral in~Eq.~\eqref{eq:(1)} and
rewrite the fermionic Lagrangian
as~$\mathcal{L}=\Bar{\psi}i\gamma^{\mu}(\partial_{\mu}-i\partial_{\mu}\alpha(x)
-i\gamma_{5}\partial_{\mu}\beta(x))\psi$. The vector part
$\partial_{\mu}\alpha(x)$ is transformed away by a suitable gauge
transformation without generating any Jacobian factor. The axial-vector part
$\partial_{\mu}\beta(x)$ is also transformed away by a suitable chiral gauge
transformation but with a non-trivial Jacobian factor. The integrated form of
the Jacobian is evaluated by the repeated applications of infinitesimal chiral
transformations
in the form~$e^{-(i/2\pi)\int d^{2}x\,\partial_{\mu}\beta(x)\partial^{\mu}\beta(x)}$, and the
final result is expressed in terms of~$v_{\mu}(x)$ as an explicit
generating functional. The final result is
\begin{equation}
   W(v_{\mu})
   =\frac{1}{2\pi}
   \int d^{2}x\,\epsilon^{\mu\nu}\partial_{\mu}v_{\nu}(x)
   \frac{1}{\partial_{\mu}\partial^{\mu}}
   \epsilon^{\alpha\beta}\partial_{\alpha}v_{\beta}(x),
\label{eq:(4)}
\end{equation}
which is non-local with respect to $v_{\mu}(x)$, and it is also derived from the bosonic path integral in~Eq.~\eqref{eq:(1)}.
See~Refs.~\cite{Roskies:1980jh,GamboaSaravi:1981zd,Furuya:1982fh,Naon:1984zp}
and the monograph~\cite{Fujikawa:2004cx} for further details.

The relation~\eqref{eq:(1)} shows that the theory of a free Dirac
fermion~$\psi$ and the theory of a free real Bose field $\xi$ define the
identical generating functional~$W(v_{\mu})$ of connected Green's functions.
The basic bosonization rule is thus
\begin{equation}
   \Bar{\psi}(x)\gamma^{\mu}\psi(x)=-\epsilon^{\mu\nu}\partial_{\nu}\xi(x),\qquad
   \text{or}\qquad
   \Bar{\psi}(x)\gamma^{\mu}\gamma_{5}\psi(x)=\partial^{\mu}\xi(x).
\label{eq:(5)}
\end{equation}
We used the relation $\gamma^{\mu}\gamma_{5}=-\epsilon^{\mu\nu}\gamma_{\nu}$
with~$\epsilon^{10}=1$, and $\gamma^{0}=\sigma^{1}$, $\gamma^{1}=i\sigma^{2}$,
$\gamma_{5}=\gamma^{1}\gamma^{0}=\sigma^{3}$. The notational convention is
summarized in~Appendix~\ref{app:A}. This current bosonization shows that the
scalar field $\xi(x)$ is a pseudo-scalar. The above path integral bosonization
(and its non-Abelian generalization) works for the known interesting examples
in~$d=2$~\cite{Fujikawa:2004cx}, for instance, the massless and massive
Thirring models (the latter corresponds to the sine-Gordon
model~\cite{Coleman:1974bu}) and the Wess--Zumino--Witten
model~\cite{Witten:1983ar}.

\section{Bosonization of anti-commuting spinors}

It is interesting to examine if the direct bosonization of~$\psi(x)$ in the
form $\psi(x)\sim\exp[i\xi(x)]$~\cite{Abdalla:1991vua,Stone:1995ys} instead of
the current in~Eq.~\eqref{eq:(5)} is possible in path integral formulation by
generalizing the formula~\eqref{eq:(1)}. For this purpose, we start with
Green's function defined by
\begin{equation}
   \Box G(x-y)=\delta^{2}(x-y),
\label{eq:(6)}
\end{equation}
which is best defined in Euclidean $d=2$ space and given by
\begin{equation}
   G_{E}(x-y)=\frac{1}{4\pi}\ln\left[(x^{1}-y^{1})^{2}+(x^{2}-y^{2})^{2}\right].
\label{eq:(7)}
\end{equation}
The corresponding Green's function in Minkowski metric is given by
\begin{equation}
   G_{M}(x-y)=\frac{-i}{4\pi}\ln\left[(x^{1}-y^{1})^{2}-(x^{0}-y^{0})^{2}\right],
\label{eq:(8)}
\end{equation}
where the imaginary factor $i$ arises from the rotation from Euclidean to
Minkowski spaces by~$x^{2}\to ix^{0}$, and thus
$\int\Box_{E}G_{E}(x^{1}-y^{1},x^{2}-y^{2})\,dx^{1}dx^{2}
=\int(-1)\Box_{M}G_{E}(x^{1}-y^{1},ix^{0}-iy^{0})i\,dx^{1}dx^{0}
=\int\Box_{M}G_{M}(x^{1}-y^{1},x^{0}-y^{0})\,dx^{1}dx^{0}=1$. This
expression~\eqref{eq:(8)} when combined with
\begin{align}
   \left(\frac{1}{2\pi}\right)
   i\gamma^{\mu}\partial_{\mu}
   \left\langle T^{\star}\psi(x)\Bar{\psi}(y)\right\rangle
   &=\int\mathcal{D}\Bar{\psi}\,\mathcal{D}\psi\,
   \left\{-i\left(\frac{\delta}{\delta\Bar{\psi}(x)}e^{iS}\right)
   \Bar{\psi}(y)\right\}
\notag\\
   &=i\delta^{2}_{M}(x-y)
\label{eq:(9)}
\end{align}
gives a correlation function of a free fermion
\begin{equation}
   \left\langle T^{\star}\psi(x)\Bar{\psi}(y)\right\rangle
   =2\pi\gamma^{\mu}\partial_{\mu}G_{M}(x-y)
   =-i\frac{\gamma^{\mu}(x-y)_{\mu}}{(x-y)^{2}}.
\label{eq:(10)}
\end{equation}
In the present paper we often use the two components of a spinor~$\psi(x)$,
which are defined in our convention summarized in~Appendix~\ref{app:A} by
\begin{equation}
   \psi(x)=\begin{pmatrix}
            \psi_{R}(x)\\
            \psi_{L}(x)\\
            \end{pmatrix},\qquad
   \Bar{\psi}(x)=\begin{pmatrix}
            \psi_{L}^{\dagger}(x)\\
            \psi_{R}^{\dagger}(x)\\
            \end{pmatrix},
\label{eq:(11)}
\end{equation}
and the correlation function~\eqref{eq:(10)} is written as
\begin{equation}
   \left\langle T^{\star}\psi_{L}(x)\psi^{\dagger}_{L}(y)\right\rangle
   =\frac{1}{i(x-y)_{+}},\qquad
   \left\langle T^{\star}\psi_{R}(x)\psi_{R}^{\dagger}(y)\right\rangle
   =\frac{1}{i(x-y)_{-}}.
\label{eq:(12)}
\end{equation}
We note that $x_{\pm}=x_{0}\pm x_{1}$ and $x^{2}=x_{+}x_{-}$ in this convention,
and the free Dirac action is written as
\begin{equation}
   S=\int d^{2}x\,\Bar{\psi}i\gamma^{\mu}\partial_{\mu}\psi
   =\int d^{2}x\,\left[\psi_{R}^{\dagger}i\partial_{-}\psi_{R}
   +\psi_{L}^{\dagger}i\partial_{+}\psi_{L}\right].
\label{eq:(13)}
\end{equation}

As for the bosonic variable $\xi(x)$, we have
\begin{align}
   \left(\frac{1}{4\pi}\right)
   \partial_{\mu}\partial^{\mu}
   \left\langle T^{\star} \xi(x)\xi(y)\right\rangle
   &=\int\mathcal{D}\xi\,
   \left(\frac{1}{4\pi}\right)\partial_{\mu}\partial^{\mu}\xi(x)\xi(y)
   \exp\left\{i\int d^{2}x\,\left[
   \left(\frac{1}{4\pi}\right)
   \frac{1}{2}\partial^{\mu}\xi\partial_{\mu}\xi\right]\right\}
\notag\\
   &=\int\mathcal{D}\xi\,
   \left[i\frac{\delta e^{iS}}{\delta \xi(x)}\right]\xi(y)
\notag\\
   &=-i\delta_{M}^{2}(x-y)
\label{eq:(14)}
\end{align}
which implies
\begin{equation}
   \left\langle T^{\star}\xi(x)\xi(y)\right\rangle
   =-i4\pi G_{M}(x-y)
   =-\ln\left[(x^{1}-y^{1})^{2}-(x^{0}-y^{0})^{2}\right],
\label{eq:(15)}
\end{equation}
and thus
\begin{align}
   \left\langle T^{\star}\exp\left[i\xi(x)\right]
   \exp\left[-i\xi(y)\right]\right\rangle
   &=\Lambda^{-2}\exp\left\{-\ln\left[-(x-y)^{2}\right]\right\}
\notag\\
   &=\Lambda^{-2}\frac{1}{\left[-(x-y)_{+}(x-y)_{-}\right]},
\label{eq:(16)}
\end{align}
where we denote the divergence at coincident points by $\Lambda$. This last
correlation function or more generally the correlation function in the presence
of the composite operator $\exp[i\xi(x)]\exp[-i\xi(y)]$ is evaluated in path
integral by adding a source term
\begin{align}
   &\int\mathcal{D}\xi\,
   \exp\left\{i\int d^{2}x\,\left[
   \left(\frac{1}{4\pi}\right)
   \frac{1}{2}\partial^{\mu}\xi\partial_{\mu}\xi+\xi(x)J(x)\right]\right\}
\notag\\
   &=\int\mathcal{D}\xi\,
   \exp\left\{i\int d^{2}x\,\left[
   \left(\frac{1}{4\pi}\right)
   \frac{1}{2}\partial^{\mu}\xi\partial_{\mu}\xi\right]\right\}
   \exp\left[\frac{i}{2}\int d^{2}u\,d^{2}v\,J(u)4\pi G_{M}(u-v)J(v)\right]
\label{eq:(17)}
\end{align}
and replacing $J(u)\to J(u)+\delta(u-x)-\delta(u-y)$. We thus have
\begin{equation}
   T^{\star}\Lambda^{2}\exp\left[i\xi(x)\right]\exp\left[-i\xi(y)\right]
   =\Lambda^{2}\left\langle T^{\star}
   \exp\left[i\xi(x)\right]\exp\left[-i\xi(y)\right]\right\rangle
   :\exp\left[i\xi(x)-i\xi(y)\right]:,
\label{eq:(18)}
\end{equation}
where normal ordering means no more contraction between $\xi(x)$ and~$\xi(y)$,
and those variables are contracted with variables contained in~$\xi(x)J(x)$
outside the composite operator in path integral.

We now recall an important formula that is shown using Cauchy's lemma, of which
proof is given in~Appendix~\ref{app:B} for completeness,
\begin{align}
   &\int\mathcal{D}\Bar{\psi}\,\mathcal{D}\psi\,
   \exp\left\{i\left(\frac{1}{2\pi}\right)
   \int d^{2}x\,\left[
   \Bar{\psi}i\gamma^{\mu}\partial_{\mu}\psi\right]\right\}
   \prod_{j=1}^{N}\prod_{k=1}^{N}\psi_{L}(x_{j})\psi_{R}^{\dagger}(x_{j})
   \psi_{R}(y_{k})\psi_{L}^{\dagger}(y_{k})
\notag\\
   &=\det\frac{1}{i(x_i-y_k)_+}\det\frac{1}{i(x_i-y_k)_-}
\notag\\
   &=\prod_{j=1}^{N}\prod_{k=1}^{N}\frac{1}{\left[-(x_{j}-y_{k})^{2}\right]}
   \prod^{N}_{j_{1}>j_{2}}\prod^{N}_{k_{1}>k_{2}}
   \left[-(x_{j_{1}}-x_{j_{2}})^{2}\right]\left[-(y_{k_{1}}-y_{k_{2}})^{2}\right]
\notag\\
   &=\int\mathcal{D}\xi\,
   \exp\left\{i\left(\frac{1}{4\pi}\right)
   \int d^{2}x\,
   \left[\frac{1}{2}\partial^{\mu}\xi(x)\partial_{\mu}\xi(x)\right]\right\}
   \prod_{j=1}^{N}\prod_{k=1}^{N}
   \Lambda\exp\left[i\xi(x_{j})\right]\Lambda\exp\left[-i\xi(y_{k})\right].
\label{eq:(19)}
\end{align}
The importance of Cauchy's lemma in bosonization has been noted
in~Ref.~\cite{Stone:1995ys}. 
We then establish
\begin{align}
   &e^{iW(v_{\mu},j_{R},j_{L})}
\notag\\
   &=\int\mathcal{D}\Bar{\psi}\,\mathcal{D}\psi\,
   \exp\left\{
   i\left(\frac{1}{2\pi}\right)
   \int d^{2}x\,\left[
   \Bar{\psi}i\gamma^{\mu}\partial_{\mu}\psi
   +v_{\mu}\Bar{\psi}\gamma^{\mu}\psi
   -j_{L}(x)\psi_{L}\psi^{\dagger}_{R}-j_{R}(x)\psi_{R}\psi^{\dagger}_{L}
   \right]\right\}
\notag\\
   &=\int\mathcal{D}\Bar{\psi}\,\mathcal{D}\psi\,
   \exp\biggl\{
   i\left(\frac{1}{2\pi}\right)
   \int d^{2}x\,
\notag\\
   &\qquad\qquad\qquad\qquad{}\times
   \left[
   \Bar{\psi}i\gamma^{\mu}\partial_{\mu}\psi
   -j_{L}(x)e^{-2i\beta}\psi_{L}\psi^{\dagger}_{R}
   -j_{R}(x)e^{2i\beta}\psi_{R}\psi^{\dagger}_{L}
   -\partial_{\mu}\beta\partial^{\mu}\beta
   \right]\biggr\}
\notag\\
   &=\int\mathcal{D}\xi\,
   \exp\biggl\{
   i\left(\frac{1}{4\pi}\right)
   \int d^{2}x\,
\notag\\
   &\qquad\qquad\qquad{}\times
   \left[
   \frac{1}{2}\partial^{\mu}\xi(x)\partial_{\mu}\xi(x)
   -2j_{L}(x)\Lambda e^{-2i\beta+i\xi}
   -2j_{R}(x)\Lambda e^{2i\beta-i\xi}
   -2\partial_{\mu}\beta\partial^{\mu}\beta
   \right]\biggr\}
\notag\\
   &=\int\mathcal{D}\xi\,
   \exp\left\{
   i\left(\frac{1}{4\pi}\right)
   \int d^{2}x\,
   \left[
   \frac{1}{2}\partial^{\mu}\xi(x)\partial_{\mu}\xi(x)
   -2v_{\mu}\epsilon^{\mu\nu}\partial_{\nu}\xi(x)
   -2j_{L}\Lambda e^{i\xi}
   -2j_{R}\Lambda e^{-i\xi}
   \right]\right\},
\label{eq:(20)}
\end{align}
where in the second line, we replaced
$v_{\mu}=\partial_{\mu}\alpha(x)+\epsilon_{\mu\nu}\partial^{\nu}\beta(x)$ and the
fermion Lagrangian is changed
to~$\mathcal{L}=\Bar{\psi}i\gamma^{\mu}
(\partial_{\mu}-i\partial_{\mu}\alpha(x)-i\gamma_{5}\partial_{\mu}\beta(x))\psi$.
The vector freedom $\partial_{\mu}\alpha(x)$ is gauge transformed away without
modifying the rest of the terms and without any Jacobian. The axial vector
freedom $\partial_{\mu}\beta(x)$ is transformed away by the change of the
fermion variables $\psi\to e^{i\beta\gamma_{5}}\psi$
and~$\Bar{\psi}\to\Bar{\psi}e^{i\beta\gamma_{5}}$, which give rise to an
integrated Jacobian factor~$-i/(2\pi)\int d^{2}x\,%
\partial_{\mu}\beta\partial^{\mu}\beta$~\cite{Roskies:1980jh,%
GamboaSaravi:1981zd,Furuya:1982fh,Naon:1984zp} together with the modifications
of the rest of the terms; this integrated Jacobian is confirmed to be correct
by considering an infinitesimal variation of~$\beta$ and obtaining
$(i/\pi)\int d^{2}x\,\delta\beta\partial_{\mu}\partial^{\mu}\beta$.

We then expand the fermionic path integral in powers of~$j_{L}$ and~$j_{R}$ in
the second line in~Eq.~\eqref{eq:(20)}, which is then converted to a bosonic
path integral using the formula~\eqref{eq:(19)}. The final result is then
re-summed to an exponential form in bosonic path integral. In the last step
in~Eq.~\eqref{eq:(20)}, we changed the path integral variable
$\xi-2\beta\to\xi$ by noting $\mathcal{D}(\xi-2\beta)=\mathcal{D}\xi$ and used
the relation~$\partial_{\nu}\partial^{\nu}\beta
=-\partial_{\nu}v_{\mu}\epsilon^{\mu\nu}$.

We have thus established path integral bosonization rules
in~Eq.~\eqref{eq:(20)},
\begin{equation}
   \Bar{\psi}\gamma^{\mu}\gamma_{5}\psi=\partial^{\mu}\xi,\qquad
   \psi_{L}(x)\psi_{R}^{\dagger}(x)=\Lambda\exp\left[i\xi(x)\right],\qquad
   \psi_{R}(x)\psi_{L}^{\dagger}(x)=\Lambda\exp\left[-i\xi(x)\right],
\label{eq:(21)}
\end{equation}
where the first relation is also written as
\begin{equation}
   \psi_{L}^{\dagger}(x)\psi_{L}(x)=-\partial^{+}\xi(x),\qquad
   \psi_{R}^{\dagger}(x)\psi_{R}(x)=\partial^{-}\xi(x).
\label{eq:(22)}
\end{equation}
The consistency of the last two relations in~Eq.~\eqref{eq:(21)} with the first
relation is confirmed by comparing (the short distance expansions)
\begin{align}
   &T^{\star}\psi_{L}(x)\psi_{R}^{\dagger}(x)\psi_{R}(y)\psi_{L}^{\dagger}(y)
\notag\\
   &=\frac{1}{i(x-y)_{+}}\frac{1}{i(x-y)_{-}}
   +\frac{1}{i(x-y)_{+}}\psi_{R}^{\dagger}(y)\psi_{R}(y)
   -\frac{1}{i(x-y)_{-}}\psi_{L}^{\dagger}(y)\psi_{L}(y)
   +\text{regular terms},
\label{eq:(23)}
\end{align}
which is established in the path integral,
\begin{align}
   &\int\mathcal{D}\Bar{\psi}\,\mathcal{D}\psi\,
   \psi_{L}(x)\psi_{R}^{\dagger}(x)\psi_{R}(y)\psi_{L}^{\dagger}(y)
   \exp\left\{
   i\left(\frac{1}{2\pi}\right)
   \int d^{2}x\,\left[
   \Bar{\psi}i\gamma^{\mu}\partial_{\mu}\psi\right]\right\}
\notag\\
   &\qquad{}
   \times\exp\left\{
   \int d^{2}x\,\left[
   \eta_{L}^{\dagger}(x)\psi_{L}(x)+\psi_{L}^{\dagger}(y)\eta_{L}(x)
   +\eta^{\dagger}_{R}(x)\psi_{R}(x)+\psi^{\dagger}_{R}(x)\eta_{R}(x)
   \right]\right\},
\label{eq:(24)}
\end{align}
by expanding in powers of source functions~$\eta(x)$, with the bosonic relation
derived from~Eq.~\eqref{eq:(18)},
\begin{align}
   &T^{\star}\Lambda^{2}\exp\left[i\xi(x)\right]\exp\left[-i\xi(y)\right]
\notag\\
   &=\frac{1}{i(x-y)_{+}}\frac{1}{i(x-y)_{-}}
   \left\{1+i(x-y)_{+}\partial^{+}\xi(y)+i(x-y)_{-}\partial^{-}\xi(y)\right\}
   +\text{regular terms},
\label{eq:(25)}
\end{align}
where regular terms mean no singularities for~$O((x-y)_{+})=O((x-y)_{-})$. These
two expansions \eqref{eq:(23)} and~\eqref{eq:(25)} are consistent with the
relations in~Eq.~\eqref{eq:(22)}.

These bosonization rules~\eqref{eq:(21)} are consistent with the parity
transformation $\psi^{\prime}(x^{\prime})=\gamma^{0}\psi(x)$
and~$\xi^{\prime}(x^{\prime})=-\xi(x)$ with~$x^{\prime}=(x^{0},-x^{1})$, and
Lorentz properties
\begin{align}
   i\Bar{\psi}(x)\gamma_{5}\psi(x)
   &=i\psi_{L}^{\dagger}(x)\psi_{R}(x)-i\psi_{R}^{\dagger}(x)\psi_{L}(x)
   =-2\Lambda \sin\xi(x),
\notag\\
   \Bar{\psi}(x)\psi(x)
   &=\psi_{L}^{\dagger}(x)\psi_{R}(x)+\psi_{R}^{\dagger}(x)\psi_{L}(x)
   =-2\Lambda\cos\xi(x).
\label{eq:(26)}
\end{align}
They are also consistent with chiral Ward-Takahashi identities in path integral
\begin{align}
   &i\partial_{\mu}
   \left\langle T^{\star}\left(\frac{1}{2\pi}\right)
   \Bar{\psi}(x)\gamma^{\mu}\gamma_{5}\psi(x)
   \prod_{j=1}^{N}\prod_{k=1}^{N}
   \psi_{L}(x_{j})\psi_{R}^{\dagger}(x_{j})
   \psi_{R}(y_{k})\psi_{L}^{\dagger}(y_{k})\right\rangle
\notag\\
   &=\sum_{j}2i\delta^{2}(x-x_{j})
   \left\langle T^{\star}\prod_{j=1}^{N}\prod_{k=1}^{N}
   \psi_{L}(x_{j})\psi_{R}^{\dagger}(x_{j})
   \psi_{R}(y_{k})\psi_{L}^{\dagger}(y_{k})\right\rangle
\notag\\
   &\qquad{}
   -\sum_{k}2i\delta^{2}(x-y_{k})
   \left\langle T^{\star}\prod_{j=1}^{N}\prod_{k=1}^{N}
   \psi_{L}(x_{j})\psi_{R}^{\dagger}(x_{j})
   \psi_{R}(y_{k})\psi_{L}^{\dagger}(y_{k})\right\rangle
\label{eq:(27)}
\end{align}
and
\begin{align}
   &i\partial_{\mu}
   \left\langle T^{\star}\left(\frac{1}{2\pi}\right)
   \partial^{\mu}\xi(x)\prod_{j=1}^{N}\prod_{k=1}^{N}
   \Lambda\exp\left[i\xi(x_{j})\right]\Lambda\exp\left[-i\xi(y_{k})\right]
   \right\rangle
\notag\\
   &=\sum_{j}2i\delta^{2}(x-x_{j})
   \left\langle T^{\star}\prod_{j=1}^{N}\prod_{k=1}^{N}
   \Lambda\exp\left[i\xi(x_{j})\right]\Lambda\exp\left[-i\xi(y_{k})\right]
   \right\rangle
\notag\\
   &\qquad{}
   -\sum_{k}2i\delta^{2}(x-y_{k})
   \left\langle T^{\star}\prod_{j=1}^{N}\prod_{k=1}^{N}
   \Lambda\exp\left[i\xi(x_{j})\right]\Lambda\exp\left[-i\xi(y_{k})\right]
   \right\rangle,
\label{eq:(28)}
\end{align}
where the bosonic form of identities are obtained by considering the change of
integration variable $\xi(x)\to\xi(x)^{\prime}=\xi(x)+\alpha(x)$ in the bosonic
path integral. Equivalently, the current algebra in operator formalism
\begin{equation}
   i\left[\left(\frac{1}{2\pi}\right)
   \Bar{\psi}(x)\gamma^{0}\gamma_{5}\psi(x),
   \psi_{L}(y)\psi_{R}^{\dagger}(y)\right]\delta(x^{0}-y^{0})
   =2i\psi_{L}(y)\psi_{R}^{\dagger}(y)\delta^{2}(x-y),
\label{eq:(29)}
\end{equation}
for example, is consistently realized by
\begin{equation}
   i\left[\left(\frac{1}{2\pi}\right)
   \partial^{0}\xi(x),\Lambda e^{i\xi(y)}\right]
   \delta(x^{0}-y^{0})
   =2i\Lambda e^{i\xi(y)}\delta^{2}(x-y),
\label{eq:(30)}
\end{equation}
since the canonical conjugate of~$\xi(x)$
is~$\Pi(x)=1/(4\pi)\partial^{0}\xi(x)$.

One may thus identify the last two relations in~Eq.~\eqref{eq:(21)} as the path
integral version of the \emph{off-shell\/} direct bosonization rules. In the
limit of \emph{on-shell fields\/} with $\partial^{+}\partial^{-}\xi(x)=0$
and~$\xi(x)=\xi(x_{+})+\xi(x_{-})$, these direct bosonization rules imply
\begin{equation}
   \psi_{L}(x_{+})=\Lambda^{1/2}e^{i\xi(x_{+})}=:e^{i\xi(x_{+})}:,\qquad
   \psi_{R}^{\dagger}(x_{-})=\Lambda^{1/2}e^{i\xi(x_{-})}=:e^{i\xi(x_{-})}:,
\label{eq:(31)}
\end{equation}
which is the familiar direct bosonization formula in operator
formalism~\cite{Abdalla:1991vua,Stone:1995ys}. Starting with the second relation of~Eq.~\eqref{eq:(21)}, which is valid for general off-shell fields, we obtain this particular form of decomposition by considering a subset of  $\psi_L(x)$ and $\psi_R^\dagger(x)$ which satisfy the equations of motion $
\partial^{-}\psi_{L}(x)=0$ and $\partial^{+}\psi_R^\dagger(x)=0$ and thus depend only on~$x_+$ and $x_-$, respectively. The factor~$\Lambda^{1/2}$ is absorbed by
normal ordering. The formula~\eqref{eq:(19)} is also consistent with the
factorization into $x_{+}$ and~$x_{-}$ sectors in the last two relations
in~Eq.~\eqref{eq:(21)}. Note the notational conventions
$x_{\pm}=x_{0}\pm x_{1}$, $x^{2}=x_{+}x_{-}$, and~$\partial^{\pm}x_{\pm}=1$.

We emphasize that we cannot define the path integral in terms of the on-shell
variables $\xi(x_{+})$ and~$\xi(x_{-})$ directly, unlike the operator formalism
such as conformal field theory where $\xi(x_{+})$ and~$\xi(x_{-})$ are treated
as independent variables.

\section{Bosonization of commuting spinors}
The bosonic spinors, namely, commuting spinors appear in the quantization of
the spinning string~\cite{Polchinski:1998rr,peter}. The Pauli--Villars
regularization is also implemented in path integrals using those bosonic
fermions~\cite{Fujikawa:2004cx}.

We here discuss the bosonization of commuting spinors in~$d=2$. We can readily
establish the relation
\begin{align}
   e^{iW(v_{\mu})}&=\int\mathcal{D}\Bar{\phi}\,\mathcal{D}\phi\,
   \exp\left\{
   i\left(\frac{1}{2\pi}\right)
   \int d^{2}x\,
   \left[\Bar{\phi}i\gamma^{\mu}\partial_{\mu}\phi
   +v_{\mu}\Bar{\phi}\gamma^{\mu}\phi\right]\right\}
\notag\\
   &=\int\mathcal{D}\xi\,
   \exp\left\{i\left(\frac{1}{4\pi}\right)
   \int d^{2}x\,\left[
   -\frac{1}{2}\partial^{\mu}\xi(x)\partial_{\mu}\xi(x)
   -2v_{\mu}\epsilon^{\mu\nu}\partial_{\nu}\xi(x)\right]\right\},
\label{eq:(32)}
\end{align}
where $\phi(x)$ stands for the commuting spinor in $d=2$. The basic difference
of this expression from the previous one~\eqref{eq:(1)} is that the signature
in front of the action for the bosonic field $\xi(x)$ differs, namely, we have
$-$ instead of~$+$. This minus sign must be adopted since the Jacobian changes
signature for the commuting variables~$\phi$ from the anti-commuting Grassmann
variables~$\psi$. The boson field~$\xi(x)$ in the present case corresponds to a
\emph{negative normed\/} field.

The bosonization rule of the current
\begin{equation}
   \Bar{\phi}\gamma^{\mu}\phi=-\epsilon^{\mu\nu}\partial_{\nu}\xi(x)
\label{eq:(33)}
\end{equation}
is the same as before
\begin{equation}
   \phi_{L}^{\dagger}(x)\phi_{L}(x)=-\partial^{+}\xi(x),\qquad
   \phi_{R}^{\dagger}(x)\phi_{R}(x)=\partial^{-}\xi(x).
\label{eq:(34)}
\end{equation}
But we have
\begin{equation}
   \left\langle T^{\star}\xi(x)\xi(y)\right\rangle
   =i4\pi G_{M}(x-y)
   =\ln\left[(x^{1}-y^{1})^{2}-(x^{0}-y^{0})^{2}\right],
\label{eq:(35)}
\end{equation}
and thus
\begin{align}
   \left\langle T^{\star}\exp\left[i\xi(x)\right]\exp\left[-i\xi(y)\right]
   \right\rangle
   &=\Lambda^{2}\exp\left[i4\pi G_{M}(x-y)\right]
\notag\\
   &=\Lambda^{2}\left[-(x-y)^{2}\right],
\label{eq:(36)}
\end{align}
which cannot reproduce the fermion poles.

However, the relation
\begin{equation}
   (-1)\left\langle
   T^{\star}\phi_{L}(x)\phi_{R}^{\dagger}(x)\phi_{R}(y)\phi_{L}^{\dagger}(y)
   \right\rangle
   =\left[\frac{1}{i(x-y)_{+}}\right]\left[\frac{1}{i(x-y)_{-}}\right],
\label{eq:(37)}
\end{equation}
is established by noting
\begin{equation}
   \left\langle
   T^{\star}\phi_{R}(x)\phi_{R}^{\dagger}(y)\right\rangle
   =\frac{1}{i(x-y)_{-}},\qquad
   \left\langle
   T^{\star}\phi_{L}(x)\phi_{L}^{\dagger}(y)\right\rangle
   =\frac{1}{i(x-y)_{+}}.
\label{eq:(38)}
\end{equation}
This may be compared with
\begin{equation}
   \left\langle
   T^{\star}\exp\left[-i\xi(x)\right]
   \partial^{+}\exp\left[-i\chi(x)\right]
   \exp\left[i\xi(y)\right]\partial^{-}\exp\left[i\chi(y)\right]
   \right\rangle
   =\left[\frac{1}{i(x-y)_{+}}\right]\left[\frac{1}{i(x-y)_{-}}\right]
\label{eq:(39)}
\end{equation}
using the positive normed \emph{auxiliary} (pseudo-scalar) field $\chi(x)$,
\begin{equation}
   \left\langle
   T^{\star}\Lambda^{2}\exp\left[i\chi(x)\right]
   \exp\left[-i\chi(y)\right]
   \right\rangle
   =\frac{1}{\left[-(x-y)^{2}\right]},
\label{eq:(40)}
\end{equation}
with~$\langle T^{\star}\chi(x)\chi(y)\rangle
=-\ln[(x^{1}-y^{1})^{2}-(x^{0}-y^{0})^{2}]$; the role of~$\chi(x)$ is to provide
a factor~$1/[-(x-y)^{2}]^{2}$ without spoiling the bosonization rule of the
current in~Eq.~\eqref{eq:(33)}. Note that $\partial^{\pm}x_{\pm}=1$
and~$\partial^{\pm}x_{\mp}=0$.

We confirm that the extra field~$\chi$, which has been used in the conformal field theory approach~\cite{Polchinski:1998rr, peter} but  was introduced in a somewhat ad
hoc manner in the present study, does not spoil the basic bosonization of the
current in~Eq.~\eqref{eq:(33)}. We check this by examining the short distance
expansion, which tests algebraic properties,
\begin{align}
   &(-1)T^{\star}\phi_{L}(x)\phi_{R}^{\dagger}(x)\phi_{R}(y)\phi_{L}^{\dagger}(y)
\notag\\
   &=\left[\frac{1}{i(x-y)_{+}}\right]
   \left[\frac{1}{i(x-y)_{-}}\right]
   +\left[\frac{1}{i(x-y)_{-}}\right]\phi_{L}(y)\phi_{L}^{\dagger}(y)
   -\left[\frac{1}{i(x-y)_{+}}\right]\phi_{R}(y)\phi_{R}^{\dagger}(y)
\notag\\
   &\qquad{}
   +\text{regular terms},
\label{eq:(41)}
\end{align}
which may be compared with (by using a relation analogous
to~Eq.~\eqref{eq:(18)})
\begin{align}
   &T^{\star}\exp\left[-i\xi(x)\right]
   \partial^{+}\exp\left[-i\chi(x)\right]
   \exp\left[i\xi(y)\right]\partial^{-}\exp\left[i\chi(y)\right]
\notag\\
   &=\frac{1}{\left[-(x-y)_{+}(x-y)_{-}\right]}
   \left\{
   1-i(x-y)_{+}\partial^{+}\xi(y)-i(x-y)_{-}\partial^{-}\xi(y)+\dotsb
   \right\}
\notag\\
   &=\left[\frac{1}{i(x-y)_{+}}\right]\left[\frac{1}{i(x-y)_{-}}\right]
   -\left[\frac{1}{i(x-y)_{-}}\right]\partial^{+}\xi(y)
   -\left[\frac{1}{i(x-y)_{+}}\right]\partial^{-}\xi(y)
\notag\\
   &\qquad{}+
   \text{regular terms},
\label{eq:(42)}
\end{align}
where regular terms mean no singularities for~$O((x-y)_{+})=O((x-y)_{-})$. These
expansions \eqref{eq:(41)} and~\eqref{eq:(42)} are consistent
with~Eq.~\eqref{eq:(34)}.

We thus expect the off-shell bosonization rules
\begin{equation}
   \phi_{L}(x)\phi^{\dagger}_{R}(x)=ie^{-i\xi(x)}\partial^{+}e^{-i\chi(x)},\qquad
   \phi_{R}(x)\phi^{\dagger}_{L}(x)=ie^{i\xi(x)}\partial^{-}e^{i\chi(x)},
\label{eq:(43)}
\end{equation}
which are consistent with parity transformation. We can in fact establish these
rules together with~Eq.~\eqref{eq:(33)} by repeating the same steps as
in~Eq.~\eqref{eq:(20)},
\begin{align}
   &e^{iW(v_{\mu},j_{R},j_{L})}
\notag\\
   &=\int\mathcal{D}\Bar{\phi}\,\mathcal{D}\phi\,
   \exp\left\{
   i\left(\frac{1}{2\pi}\right)
   \int d^{2}x\,\left[
   \Bar{\phi}i\gamma^{\mu}\partial_{\mu}\phi
   +v_{\mu}(x)\Bar{\phi}\gamma^{\mu}\phi
   -j_{L}(x)\phi_{L}\phi^{\dagger}_{R}
   -j_{R}(x)\phi_{R}\phi^{\dagger}_{L}\right]\right\}
\notag\\
   &=\int\mathcal{D}\Bar{\phi}\,\mathcal{D}\phi\,
   \exp\biggl\{
   i\left(\frac{1}{2\pi}\right)
   \int d^{2}x\,
\notag\\
   &\qquad\qquad\qquad\qquad\qquad{}
   \times
   \left[
   \Bar{\phi}i\gamma^{\mu}\partial_{\mu}\phi
   -j_{L}(x)e^{-2i\beta}\phi_{L}\phi^{\dagger}_{R}
   -j_{R}(x)e^{2i\beta}\phi_{R}\phi^{\dagger}_{L}
   +\partial_{\mu}\beta\partial^{\mu}\beta\right]\biggr\}
\notag\\
   &=\int\mathcal{D}\xi\,\mathcal{D}\chi\,
   \exp\biggl\{
   i\left(\frac{1}{4\pi}\right)
   \int d^{2}x\,
   \biggl[-\frac{1}{2}\partial^{\mu}\xi(x)\partial_{\mu}\xi(x)
   +\frac{1}{2}\partial^{\mu}\chi(x)\partial_{\mu}\chi(x)
\notag\\
   &\qquad\qquad\qquad\qquad\qquad{}
   -2j_{L}(x) ie^{-2i\beta-i\xi}\partial^{+}e^{-i\chi}
   -2j_{R}(x) ie^{2i\beta+i\xi}\partial^{-}e^{i\chi}
   +2\partial_{\mu}\beta\partial^{\mu}\beta
   \biggr]\biggr\}
\notag\\
   &=\int\mathcal{D}\xi\,\mathcal{D}\chi\,
   \exp\biggl\{
   i\left(\frac{1}{4\pi}\right)\int d^{2}x\,
   \biggl[-\frac{1}{2}\partial^{\mu}\xi(x)\partial_{\mu}\xi(x)
   +\frac{1}{2}\partial^{\mu}\chi(x)\partial_{\mu}\chi(x)
   -2v_{\mu}(x)\epsilon^{\mu\nu}\partial_{\nu}\xi(x)
\notag\\
   &\qquad\qquad\qquad\qquad\qquad\qquad\qquad\qquad{}
   -2j_{L}(x) ie^{-i\xi}\partial^{+}e^{-i\chi}
   -2j_{R}(x) ie^{i\xi}\partial^{-}e^{i\chi}\biggr]\biggr\},
\label{eq:(44)}
\end{align}
where in the second line, we replaced
$v_{\mu}=\partial_{\mu}\alpha(x)+\epsilon_{\mu\nu}\partial^{\nu}\beta(x)$ and
eliminated $\partial_{\mu}\alpha$ and~$\partial^{\nu}\beta$ by suitable gauge
transformations. Note the sign change of the Jacobian
factor~$\partial_{\mu}\beta\partial^{\mu}\beta$. In the second line, we used a
bosonic-spinor analogue of Cauchy's lemma in~Eq.~\eqref{eq:(19)},
\begin{align}
   &\int\mathcal{D}\Bar{\phi}\,\mathcal{D}\phi\,
   \exp\left\{
   i\left(\frac{1}{2\pi}\right)
   \int d^{2}x\,
   \left[\Bar{\phi}i\gamma^{\mu}\partial_{\mu}\phi\right]\right\}
   \prod_{j=1}^{N}\prod_{k=1}^{N}
   \phi_{L}(x_{j})\phi_{R}^{\dagger}(x_{j})\phi_{R}(y_{k})\phi_{L}^{\dagger}(y_{k})
\notag\\
   &=\left[\sum_{\{k_{1},\dots,k_{N}\}}
   \frac{1}{i(x_{1}-y_{k_{1}})_{+}}\frac{1}{i(x_{2}-y_{k_{2}})_{+}}\dotsm
   \frac{1}{i(x_{N}-y_{k_{N}})_{+}}\right]
\notag\\
   &\qquad\qquad\qquad\qquad\qquad{}
   \times
   \left[\sum_{\{k_{1},\dots,k_{N}\}}
   \frac{1}{i(y_{1}-x_{k_{1}})_{-}}\frac{1}{i(y_{2}-x_{k_{2}})_{-}}\dotsm
   \frac{1}{i(y_{N}-x_{k_{N}})_{-}}\right]
\notag\\
   &=\int\mathcal{D}\xi\,\mathcal{D}\chi\,
   \exp\left\{
   i\left(\frac{1}{4\pi}\right)
   \int d^{2}x\,
   \left[
   -\frac{1}{2}\partial^{\mu}\xi(x)\partial_{\mu}\xi(x)
   +\frac{1}{2}\partial^{\mu}\chi(x)\partial_{\mu}\chi(x)\right]\right\}
\notag\\
   &\qquad\qquad\qquad\qquad\qquad\qquad\qquad{}
   \times\prod_{j=1}^{N}\prod_{k=1}^{N}
   \left[ie^{-i\xi(x_{j})}\partial^{+}e^{-i\chi(x_{j})}\right]
   \left[ie^{i\xi(y_{k})}\partial^{-}e^{i\chi(y_{k})}\right],
\label{eq:(45)}
\end{align}
where the summation runs over all the permutations of $(1,2,\dots,N)$. The
proof of this formula is given in~Appendix~\ref{app:C}. In the last step
in~Eq.~\eqref{eq:(44)}, we changed the integration variable $\xi+2\beta\to\xi$.

For \emph{on-shell fields\/} with~$\xi(x)=\xi(x_{+})+\xi(x_{-})$
and~$\chi(x)=\chi(x_{+})+\chi(x_{-})$, the relations in~Eq.~\eqref{eq:(43)}
imply
\begin{align}
   \phi_{L}(x_{+})&=ie^{-i\xi(x_{+})}\partial^{+}e^{-i\chi(x_{+})},&
   \phi^{\dagger}_{R}(x_{-})&=e^{-i\xi(x_{-})-i\chi(x_{-})},
\notag\\
   \phi_{R}(x_{-})&=ie^{i\xi(x_{-})}\partial^{-}e^{+i\chi(x_{-})},&
   \phi^{\dagger}_{L}(x_{+})&=e^{i\xi(x_{+})+i\chi(x_{+})}.
\label{eq:(46)}
\end{align}
Here, we have noted that $\phi_L(x)$
and~$\phi_R^\dagger(x)$ depend only on~$x_+$ and~$x_-$, respectively, if one considers a subset of $\phi_L(x)$
and~$\phi_R^\dagger(x)$ which satisfy the equations of motion $
\partial^{-}\phi_{L}(x)=0$ and $\partial^{+}\phi_R^\dagger(x)=0$. The parity
symmetry is preserved in these formulas, but the hermiticity such as
$(\phi_{L}(x_{+}))^{\dagger}=\phi^{\dagger}_{L}(x_{+})$ is lost. In Euclidean
formulation such as in conformal field theory, one may treat $\phi_{L}(x_{+})$
and~$\phi^{\dagger}_{L}(x_{+})$ as independent quantities and thus this
formulation may be used; in fact, this trick has been used for the commuting
ghosts in spinning string theory where $\phi_{L}(x_+)\to-i\beta$
and~$\phi^{\dagger}_{L}(x_{+})\to\gamma$, and thus these two fields are literally
independent~\cite{Polchinski:1998rr,peter}. To be precise, after the Wick
rotation $x_{+}\to-z$ and~$x_{-}\to\Bar{z}$, we find the correspondence (after a
suitable re-definition of field variables),
\begin{align}
   \beta(z)&=e^{-\phi(z)+\chi(z)}\partial_{z}\chi(z),&
   \gamma(z)&=e^{\phi(z)-\chi(z)},
\notag
\\
  \Bar{\beta}(\Bar{z})
   &=e^{-\Bar{\phi}(\Bar{z})+\Bar{\chi}(\Bar{z})}\partial_{\Bar{z}}
   \Bar{\chi}(\Bar{z}),&
   \Bar{\gamma}(\Bar{z})&=e^{\Bar{\phi}(\Bar{z})-\Bar{\chi}(\Bar{z})}.
\label{eq:(47)}
\end{align}
See, for example, Section~10.4 in~Ref.~\cite{Polchinski:1998rr}.

\section{Discussion}
We have shown that the direct on-shell bosonization such
as~$\psi_{L}(x_{+})=e^{i\xi(x_{+})}$ and~$\psi_{R}^{\dagger}(x_{-})=e^{i\xi(x_{-})}$
in~$d=2$ is understood in a systematic way by path integral formulation which
is based on off-shell fields. This fact must be conceptually satisfactory when
one considers the path integral bosonization. The bosonization itself has many
applications in mathematical physics such as conformal field theory and also in
condensed matter physics.

Finally, we mention the first quantization of string theory briefly, where the
bosonization of Faddeev-Popov ghosts historically played important
roles~\cite{Friedan:1985ge}. For bosonic critical string theory, the
anti-commuting reparameterization ghosts are described by a formal Dirac
Lagrangian in~$d=2$ Euclidean space~\cite{Fujikawa:1981yy}
\begin{equation}
   \int\mathcal{D}\xi\,\mathcal{D}\eta\,
   \exp\left\{
   \left(\frac{1}{2\pi}\right)
   \int d^{2}x\,\xi(x)
   \left(\sigma_{1}\partial_{1}+\sigma_{3}\partial_{2}\right)\eta(x)\right\},
\label{eq:(48)}
\end{equation}
where
\begin{align}
   \xi(x)=\begin{pmatrix}
            \xi_{1}\\
            \xi_{2}\\
            \end{pmatrix},\qquad
   \eta(x)=\begin{pmatrix}
            \eta^{1}\\
            \eta^{2}\\
            \end{pmatrix},
\label{eq:(49)}
\end{align}
and $\eta$~stands for the reparameterization ghosts and $\xi(x)$ for the
multiplier fields. Our bosonization of anti-commuting spinors is directly
applied to this case, and the result gives rise to the formula in the modern
notation of $bc$ ghosts~\cite{Polchinski:1998rr,peter}. See
also~Refs.\cite{Polyakov:1981rd,Alvarez:1982zi}. For spinning critical string
theory, the path integral for the commuting ghosts associated with the fixing
of supersymmetry transformation of gravitino is described by a formal Dirac
Lagrangian in $d=2$ Euclidean space~\cite{Bouwknegt:1985rz,Rocek:1986iz}
\begin{equation}
   \int\mathcal{D}\Bar{C}\,\mathcal{D}C\,
   \exp\left\{\frac{1}{2\pi}
   \int d^{2}x\,\Bar{C}(x)
   \left(\sigma_{1}\partial_{1}-\sigma_{2}\partial_{2}\right)C(x)\right\},
\label{eq:(50)}
\end{equation}
where $C(x)$ stands for commuting spinor ghosts and $\Bar{C}(x)$ for the
multiplier fields. This is bosonized using the above result of commuting
spinors, and the result agrees with the bosonization in the modern notation
of~$\beta\gamma$ ghosts~\cite{Polchinski:1998rr,peter}. In these primitive
notations \eqref{eq:(48)} and~\eqref{eq:(50)}, the bosonization of two
component spinors appears in a more explicit manner.

\appendix
\section{Notational convention}
\label{app:A}

We here summarize the notations in~$2$-dimensional Minkowski space-time which
is defined by
\begin{equation}
   \left\{\gamma^{\mu},\gamma^{\nu}\right\}=2g^{\mu\nu}
\label{eq:(51)}
\end{equation}
with~$g_{\mu\nu}=(1,-1)=g^{\mu\nu}$ and~$\epsilon^{10}=1$. We define
$\gamma^{0}=\sigma^{1}$, $\gamma^{1}=i\sigma^{2}$,
$\gamma_{5}=\gamma^{1}\gamma^{0}=\sigma^{3}$
with~$\gamma^{\mu}\gamma_{5}=-\epsilon^{\mu\nu}\gamma_{\nu}$. The following
relations are convenient
\begin{align}
   &\gamma^{0}\frac{(1-\gamma_{5})}{2}
   =\frac{\gamma^{0}+\gamma^{1}}{2}
   \equiv\gamma^{+},\qquad
   \gamma^{1}\frac{(1-\gamma_{5})}{2}
   =\frac{\gamma^{1}+\gamma^{0}}{2}
   =\gamma^{+},
\notag\\
   &V_{\mu}\gamma^{\mu}\frac{(1-\gamma_{5})}{2}
   =(V_{0}+V_{1})(\frac{\gamma^{0}+\gamma^{1}}{2})
   \equiv V_{+}\gamma^{+}.
\label{eq:(52)}
\end{align}
Similarly
\begin{align}
   &\gamma^{0}\frac{(1+\gamma_{5})}{2}
   =\frac{\gamma^{0}-\gamma^{1}}{2}
   \equiv\gamma^{-},\qquad
   \gamma^{1}\frac{(1+\gamma_{5})}{2}
   =\frac{\gamma^{1}-\gamma^{0}}{2}
   =-\gamma^{-},
\notag\\
   &V_{\mu}\gamma^{\mu}\frac{(1+\gamma_{5})}{2}
   =(V_{0}-V_{1})(\frac{\gamma^{0}-\gamma^{1}}{2})
   \equiv V_{-}\gamma^{-},
\notag\\
   &A_{\pm}=A_{0}\pm A_{1},\qquad
   A^{\pm}=\frac{1}{2}(A^{0}\pm A^{1}),\qquad
   x_{\pm}=x_{0}\pm x_{1},
\notag\\
   &\partial_{\pm}=\partial_{0}\pm \partial_{1},\qquad
   \partial^{\pm}=\frac{1}{2}(\partial^{0}\pm\partial^{1})
   =\frac{1}{2}(\partial_{0}\mp\partial_{1})=\frac{1}{2}\partial_{\mp},
\notag\\
   &A_{\mu}B^{\mu}=\eta^{+-}A_{+}B_{-}+\eta^{-+}A_{-}B_{+},\qquad
   \eta^{+-}=\eta^{-+}=\frac{1}{2},
\notag\\
   &\epsilon^{\mu\nu}A_{\mu}B_{\nu}
   =\epsilon^{+-}A_{+}B_{-}+\epsilon^{-+}A_{-}B_{+},\qquad
   \epsilon^{+-}=-\epsilon^{-+}=\frac{1}{2}.
\label{eq:(53)}
\end{align}

\section{Proof of Cauchy's lemma}
\label{app:B}

We set
\begin{equation}
   f_N(x_1,x_2,\dots,x_N;y_1,y_2,\dots,y_N)
   \equiv\det_{j,k}\left(\frac{1}{x_j-y_k}\right)
\label{eq:(54)}
\end{equation}
and show that
\begin{align}
   &f_N(x_1,x_2,\dots,x_N;y_1,y_2,\dots,y_N)
\notag\\
   &=(-1)^{N(N-1)/2}
   \frac{\prod_{j_1>j_2}^N(x_{j_1}-x_{j_2})
   \prod_{k_1>k_2}^N(y_{k_1}-y_{k_2})}
   {\prod_{j=1}^N\prod_{k=1}^N(x_j-y_k)}
\label{eq:(55)}
\end{align}
by mathematical induction.

It is easy to see that the assertion is true for~$N=1$ (for this case, the
numerator of the right-hand side of~Eq.~\eqref{eq:(55)} is interpreted to be
unity) and for~$N=2$. Assuming that Eq.~\eqref{eq:(55)} is true when $N$ is
replaced by $N-1$, we then show that it holds also for~$N=N$.

It is obvious that, as a complex function of~$x_N$, $f_N$
in~Eq.~\eqref{eq:(54)} possesses simple poles at~$y_k$ ($k=1$, \dots, $N$).
This property is shared also by the right-hand side of~Eq.~\eqref{eq:(55)}.
Therefore, if the residues of both sides of~Eq.~\eqref{eq:(55)} at the
pole~$x_N=y_k$ are the same, the difference of both sides is a bounded entire
function of~$x_N$ which (obviously) vanishes as~$|x_N|\to\infty$. This implies
the equality~\eqref{eq:(55)} by Liouville's theorem.

Now, from the definition~\eqref{eq:(54)}, the residue of the left-hand side
of~Eq.~\eqref{eq:(55)} at the pole~$x_N=y_k$ is given by the minor determinant,
the determinant of the sub-matrix obtained by deleting the $N$-th row and
$k$-th column. That is,
\begin{equation}
   (-1)^{N+k}f_{N-1}(x_1,x_2,\dots,x_{N-1};y_1,\dots,y_{k-1},y_k,\dots,y_N).
\label{eq:(56)}
\end{equation}

On the other hand, the residue of the right-hand side of~Eq.~\eqref{eq:(55)} at
the pole~$x_N=y_k$ is
\begin{align}
   &(-1)^{N(N-1)/2}
   \left.
   \frac{\prod_{j_1>j_2}^N(x_{j_1}-x_{j_2})
   \prod_{k_1>k_2}^N(y_{k_1}-y_{k_2})}
   {\prod_{j=1}^{N-1}\prod_{l=1}^N(x_j-y_l)\prod_{m\neq k}^N(x_N-y_m)}
   \right|_{x_N=y_k}
\notag\\
   &=(-1)^{N(N-1)/2}
   \frac{\prod_{n=1}^{N-1}(y_k-x_n)\prod_{j_1>j_2}^{N-1}(x_{j_1}-x_{j_2})
   \prod_{k_1>k_2}^N(y_{k_1}-y_{k_2})}
   {\prod_{j=1}^{N-1}\prod_{l=1}^N(x_j-y_l)\prod_{m\neq k}^N(y_k-y_m)}
\notag\\
   &=
   (-1)^{N(N-1)/2}(-1)^{N-1}(-1)^{N-k}
   \frac{\prod_{j_1>j_2}^{N-1}(x_{j_1}-x_{j_2})
   \prod_{k_1>k>k_2}^N(y_{k_1}-y_{k_2})}
   {\prod_{j=1}^{N-1}\prod_{l\neq k}^N(x_j-y_l)}.
\label{eq:(57)}
\end{align}
Assuming that Eq.~\eqref{eq:(55)} is valid when $N$ is replaced by~$N-1$, it
can be confirmed that this last expression in~Eq.~\eqref{eq:(57)} is equal
to~Eq.~\eqref{eq:(56)} and thus the residues of both sides
of~Eq.~\eqref{eq:(55)} are the same. By mathematical induction, this proves
Eq.~\eqref{eq:(55)} for any~$N$.

\section{Proof of an analogue of Cauchy's lemma for a bosonic spinor in Eq.~\eqref{eq:(45)}}
\label{app:C}

The rightmost expression of~Eq.~\eqref{eq:(45)} is written as
\begin{align}
   &i^Nf_N((x_1)_+,(x_2)_+,\dots,(x_N)_+;(y_1)_+,(y_2)_+,\dots,(y_N)_+)
\notag\\
   &\qquad{}
   \times
   i^Nf_N((y_1)_-,(y_2)_-,\dots,(y_N)_-;(x_1)_-,(x_2)_-,\dots,(x_N)_-),
\label{eq:(58)}
\end{align}
if one introduces the function
\begin{align}
   &f_N(x_1,x_2,\dots,x_N;y_1,y_2,\dots,y_N)
\notag\\
   &\equiv\frac{1}{C_N(x_1,x_2,\dots,x_N;y_1,y_2,\dots,y_N)}
   \left(\prod_{l=1}^N\frac{\partial}{\partial x_l}\right)
   C_N(x_1,x_2,\dots,x_N;y_1,y_2,\dots,y_N),
\label{eq:(59)}
\end{align}
where
\begin{equation}
   C_N(x_1,x_2,\dots,x_N;y_1,y_2,\dots,y_N)
   \equiv
   \prod_{j=1}^N\prod_{k=1}^N\frac{1}{x_j-y_k}\prod_{j_1>j_2}^N
   (x_{j_1}-x_{j_2}).
\label{eq:(60)}
\end{equation}

In what follows, we show that
\begin{equation}
   f_N(x_1,x_2,\dots,x_N;y_1,y_2,\dots,y_N)
   =\sum_{\{k_1,k_2,\dots, k_N\}}
   \frac{-1}{x_1-y_{k_1}}\frac{-1}{x_2-y_{k_2}}\dotsm\frac{-1}{x_N-y_{k_N}}
\label{eq:(61)}
\end{equation}
by mathematical induction. If this is true, then the rightmost term
of~Eq.~\eqref{eq:(45)} is
\begin{align}
   &\left[\sum_{\{k_1,k_2,\dots, k_N\}}
   \frac{1}{i(x_1-y_{k_1})_+}\frac{1}{i(x_2-y_{k_2})_+}
   \dotsm\frac{1}{i(x_N-y_{k_N})_+}\right]
\notag\\
   &\times
   \left[\sum_{\{k_1,k_2,\dots, k_N\}}
   \frac{1}{i(y_1-x_{k_1})_-}\frac{1}{i(y_2-x_{k_2})_-}
   \dotsm\frac{1}{i(y_N-x_{k_N})_-}\right]
\label{eq:(62)}
\end{align}
and this coincides with the path integral over bosonic spinors on the left-hand
side of~Eq.~\eqref{eq:(45)}.

It is easy to confirm that the assertion~\eqref{eq:(61)} is true for~$N=1$ (for
which the last factor in~Eq.~\eqref{eq:(60)},
$\prod_{j_1>j_2}^N(x_{j_1}-x_{j_2})$, is interpreted as unity) and for~$N=2$.
Assuming that Eq.~\eqref{eq:(61)} is true when $N$ is replaced by~$N-1$, we
then show that it holds also for~$N=N$.

From its definition in~Eqs.~\eqref{eq:(59)} and~\eqref{eq:(60)}, it is obvious
that $f_N$, as the function of~$x_N$, is a sum of simple poles possibly
at~$x_j$ ($j\neq N$) and~$y_k$. Thus, to conclude~Eq.~\eqref{eq:(61)}, it
suffices to determine the position of all poles as the function of~$x_N$ and
the residue at each pole.

Now we compute
\begin{align}
   &\frac{\partial}{\partial x_N}
   C_N(x_1,x_2,\dots,x_N;y_1,y_2,\dots,y_N)
\notag\\
   &=C_N(x_1,x_2,\dots,x_N;y_1,y_2,\dots,y_N)
   \left(\sum_{k=1}^N\frac{-1}{x_N-y_k}
   +\sum_{j=1}^{N-1}\frac{1}{x_N-x_j}\right)
\label{eq:(63)}
\end{align}
and
\begin{align}
   &\frac{\partial}{\partial x_{N-1}}\frac{\partial}{\partial x_N}
   C_N(x_1,x_2,\dots,x_N;y_1,y_2,\dots,y_N)
\notag\\
   &=C_N(x_1,x_2,\dots,x_N;y_1,y_2,\dots,y_N)
\notag\\
   &\qquad{}\times
   \Biggl[\left(
   \frac{-1}{x_N-x_{N-1}}
   +\sum_{k=1}^N\frac{-1}{x_{N-1}-y_k}
   +\sum_{j=1}^{N-2}\frac{1}{x_{N-1}-x_j}
   \right)
\notag\\
   &\qquad\qquad{}\times
   \left(
   \frac{1}{x_N-x_{N-1}}
   +\sum_{n=1}^N\frac{-1}{x_N-y_n}
   +\sum_{m=1}^{N-2}\frac{1}{x_N-x_m}
   \right)
   +\frac{1}{(x_N-x_{N-1})^2}\Biggr].
\label{eq:(64)}
\end{align}

First, examining the behavior of~Eq.~\eqref{eq:(64)} near~$x_N\sim x_{N-1}$, we
immediately find that Eq.~\eqref{eq:(64)} is regular at~$x_N=x_{N-1}$. Since
in~Eq.~\eqref{eq:(59)}, further derivatives
$\prod_{l=1}^{N-2}\partial/\partial x_l$ on~Eq.~\eqref{eq:(64)} does not produce
any singularity at~$x_N=x_{N-1}$, we conclude that~$f_N$ has no pole
at~$x_N=x_{N-1}$. Since the choice $x_{N-1}$ is of course arbitrary, we see that
$f_N$ has no pole at~$x_j$ as the function of~$x_N$.

Next, from~Eqs.~\eqref{eq:(63)} and~\eqref{eq:(59)}, we see that the residue of
the pole of~$f_N$ at~$x_N=y_N$ is given by
\begin{align}
   &(-1)
   \left[
   \prod_{m=1}^{N-1}\prod_{n=1}^N\frac{1}{x_m-y_n}\prod_{m_1>m_2}^N
   (x_{m_1}-x_{m_2})
   \right]^{-1}
\notag\\
   &\qquad\qquad\qquad\qquad{}\times
   \left.
   \left(\prod_{l=1}^{N-1}\frac{\partial}{\partial x_l}\right)
   \prod_{j=1}^{N-1}\prod_{k=1}^N\frac{1}{x_j-y_k}\prod_{j_1>j_2}^N
   (x_{j_1}-x_{j_2})
   \right|_{x_N=y_N}
\notag\\
   &=(-1)
   \left[
   \prod_{m=1}^{N-1}\prod_{n=1}^{N-1}\frac{1}{x_m-y_n}\prod_{m_1>m_2}^{N-1}
   (x_{m_1}-x_{m_2})
   \right]^{-1}
\notag\\
   &\qquad\qquad\qquad\qquad{}\times
   \left(\prod_{l=1}^{N-1}\frac{\partial}{\partial x_l}\right)
   \prod_{j=1}^{N-1}\prod_{k=1}^{N-1}\frac{1}{x_j-y_k}\prod_{j_1>j_2}^{N-1}
   (x_{j_1}-x_{j_2})
\notag\\
   &=-f_{N-1}(x_1,x_2,\dots,x_{N-1};y_1,y_2,\dots,y_{N-1}).
\label{eq:(65)}
\end{align}
This is the function~$f_N$ itself with one $N$ lower. The residue at the pole
at~$x_N=y_k$ for arbitrary~$k$ is similarly obtained. Using this information
and assuming that Eq.~\eqref{eq:(61)} with $N$ replaced by $N-1$ is valid, we
see that Eq.~\eqref{eq:(61)} is true also for~$N=N$. Eq.~\eqref{eq:(61)} holds
for any~$N$ by mathematical induction.

\section*{Acknowledgments}
We thank Peter van Nieuwenhuizen for stimulating discussions on various aspects
of quantum anomalies, which motivated the present work in particular. One of us
(K.F.) thanks the hospitality at C.~N. Yang Institute for Theoretical Physics
where the present work was initiated. We also thank Henry Tye for asking the
direct bosonization in path integral formulation.
This work is supported in part by Japan Society for the Promotion of Science
KAKENHI (Grant Nos.\ 25400415 and~23540330).

\end{document}